\def\etc{{\it etc.}}
\def\etal{{\it et al.}}
\def\ie{{\it i.e.}}

\def\~{{$\tilde{\phantom{a}}$}}

\documentclass [12pt] {article}
\usepackage{epsfig}
\usepackage{color}

\def\thebibliography#1{\section{References}\markboth
 {REFERENCES}{REFERENCES}\list
 {[\arabic{enumi}]}{\settowidth\labelwidth{[#1]}\leftmargin\labelwidth
 \advance\leftmargin\labelsep
 \usecounter{enumi}}
 \def\newblock{\hskip .11em plus .33em minus -.07em}
 \sloppy
 \sfcode`\.=1000\relax}
\def\upcite#1{\raise6pt\hbox{\scriptsize
\cite{#1}}}
\def\lsim{\mathrel {\vcenter {\baselineskip 0pt \kern 0pt
    \hbox{$<$} \kern 0pt \hbox{$\sim$} }}}
\def\gsim{\mathrel {\vcenter {\baselineskip 0pt \kern 0pt
    \hbox{$>$} \kern 0pt \hbox{$\sim$} }}}
\def\gtlt{\mathrel {\vcenter {\baselineskip 0pt \kern 0pt
    \hbox{$>$} \kern 0pt \hbox{$<$} }}}

\def\hline{\noalign{\hrule \vskip2pt}}

\def\|{\ifmmode\Vert\else \char`\|\fi}
\ifx\oldzeta\undefined                          % hasn't been done yet, so 
  \let\oldzeta=\zeta                            % save old definiton
  \def\zzeta{{\raise 2pt\hbox{$\oldzeta$}}}     % make new definition
  \let\zeta=\zzeta                              % and attatch it
\fi

\ifx\oldchi\undefined                           % hasn't been done yet, so 
  \let\oldchi=\chi                              % save old definiton
  \def\cchi{{\raise 2pt\hbox{$\oldchi$}}}       % make new definition
  \let\chi=\cchi                                % and attatch it
\fi

\def\frac#1#2{{#1 \over #2}}

\def\half{\ifinner {\scriptstyle {1 \over 2}}
   \else {1 \over 2} \fi}

\def\ave#1{\left\langle#1\right\rangle} % \ave{stuff} gives <stuff>
\def\simge{\mathrel{%
   \rlap{\raise 0.511ex \hbox{$>$}}{\lower 0.511ex \hbox{$\sim$}}}}
\def\simle{\mathrel{
   \rlap{\raise 0.511ex \hbox{$<$}}{\lower 0.511ex \hbox{$\sim$}}}}

\def\buildchar#1#2#3{{\null\!                   % \null, cancel space
   \mathop#1\limits^{#2}_{#3}                   % #1, #2 above, #3 below
   \!\null}}                                    % cancel space, \null
\def\overcirc#1{\buildchar{#1}{\circ}{}}

\def\slashchar#1{\setbox0=\hbox{$#1$}           % set a box for #1 
   \dimen0=\wd0                                 % and get its size
   \setbox1=\hbox{/} \dimen1=\wd1               % get size of /
   \ifdim\dimen0>\dimen1                        % #1 is bigger
      \rlap{\hbox to \dimen0{\hfil/\hfil}}      % so center / in box
      #1                                        % and print #1
   \else                                        % / is bigger
      \rlap{\hbox to \dimen1{\hfil$#1$\hfil}}   % so center #1
      /                                         % and print /
   \fi}                                         %

\def\subrightarrow#1{%                          % #1 under arrow
  \setbox0=\hbox{%                              % set a box
    $\displaystyle\mathop{}%                    % no mathop
    \limits_{#1}$}%                             % just limits
  \dimen0=\wd0%                                 % get width
  \advance \dimen0 by .5em%                     % add a bit
  \mathrel{%                                    % space like =
    \mathop{\hbox to \dimen0{\rightarrowfill}}% % arrow to width
       \limits_{#1}}}                           % text below

\def\overlay#1#2{\ifmmode%
\setbox0=\hbox{$#1$}%
\setbox1=\hbox to\wd0{\hss$#2$\hss}\else%
\setbox0=\hbox{#1}%
\setbox1=\hbox to\wd0{\hss#2\hss}\fi%
#1\hskip-\wd0\box1 }

\def\pmb#1{\leavevmode\setbox0=\hbox{#1}%
\kern-.02em\copy0\kern-\wd0
\kern.04em\copy0\kern-\wd0
\kern-.02em\raise.04em\box0 }

\def\vereq#1#2{\lower3pt\vbox{\baselineskip1.5pt \lineskip1.5pt
\ialign{$\m@th#1\hfill##\hfil$\crcr#2\crcr\sim\crcr}}}

\def\tensor#1{\protect\@ontopof{#1}{\leftrightarrow}{1.15}\mathord{\box2}}
\def\overstar#1{\protect\@ontopof{#1}{\ast}{1.15}\mathord{\box2}}
\def\overdots#1{\protect\@ontopof{#1}{\cdots}{1.0}\mathord{\box2}}
\def\overcirc#1{\protect\@ontopof{#1}{\circ}{1.2}\mathord{\box2}}
\def\loarrow#1{\protect\@ontopof{#1}{\leftarrow}{1.15}\mathord{\box2}}
\def\roarrow#1{\protect\@ontopof{#1}{\rightarrow}{1.15}\mathord{\box2}}

\def\@ontopof#1#2#3{%
{\mathchoice
{\@@ontopof{#1}{#2}{#3}\displaystyle\scriptstyle}%
{\@@ontopof{#1}{#2}{#3}\textstyle\scriptstyle}%
{\@@ontopof{#1}{#2}{#3}\scriptstyle\scriptscriptstyle}%
{\@@ontopof{#1}{#2}{#3}\scriptscriptstyle\scriptscriptstyle}%
}%
}

\def\@@ontopof#1#2#3#4#5{%
\setbox0=\hbox{$#4#1$}%
\setbox1=\hbox{$#5#2$}%
\setbox2=\hbox{}\ht2=\ht0 \dp2=\dp0 %
\ifdim\wd0>\wd1 %
\setbox1=\hbox to\wd0{\hss\box1\hss}%
\mathord{\rlap{\raise#3\ht0\box1}\box0}%
\else   %
\setbox1=\hbox to.9\wd1{\hss\box1\hss}%
\setbox0=\hbox to\wd1{\hss$#4\relax#1$\hss}%
\mathord{\rlap{\copy0}\raise#3\ht0\box1}%
\fi
}%

\def\lambdabar{\protect\@lambdabar}
\def\@lambdabar{%
\relax
\bgroup
\def\@tempa{\hbox{\raise.73\ht0
\hbox to0pt{\kern.25\wd0\vrule width.5\wd0
height.1pt depth.1pt\hss}\box0}}%
\mathchoice{\setbox0\hbox{$\displaystyle\lambda$}\@tempa}%
{\setbox0\hbox{$\textstyle\lambda$}\@tempa}%
{\setbox0\hbox{$\scriptstyle\lambda$}\@tempa}%
{\setbox0\hbox{$\scriptscriptstyle\lambda$}\@tempa}%
\egroup
}

\def\corresponds{{\lower.2ex\hbox{=}}{\rm\kern-.75em^\triangle}}
\def\succsim{\succ\kern-.9em_\sim\kern.3em}
\def\precsim{\prec\kern-1em_\sim\kern.3em}
\def\slantfrac#1#2{\kern1em^{#1}\kern-.3em/\kern-.1em_{#2}}

\begin{document}
                                                                
\begin{center}
{\Large\bf An Electrostatic Wave}
\\

\medskip

Kirk T.~McDonald
\\
{\sl Joseph Henry Laboratories, Princeton University, Princeton, NJ 08544}
\\
(July 28, 2002)
\end{center}

\section{Problem}

All electrostatic fields {\bf E} (\ie, ones with no time 
dependence) can be derived from a scalar potential $V$ (${\bf E} =
- \nabla V$) and hence obey $\nabla \times {\bf E} = 0$.  The latter condition
is sometimes considered to be a requirement for electrostatic fields.
Show, however, that there can exist time-dependent electric fields for
which $\nabla \times {\bf E} = 0$, which have been given the name
``electrostatic waves''.
 
In particular, show that a plane wave with electric field {\bf E} 
parallel to the wave vector {\bf k} (a longitudinal wave)
can exist in a medium with no time-dependent magnetic field if the
electric displacement {\bf D} is zero.  This cannot occur in an
ordinary dielectric medium, but can happen in a plasma.  (Time-independent
electric and magnetic fields could, of course, be superimposed on the
wave field.)\ \  Compare the potentials for the ``electrostatic wave''
in the Coulomb and Lorentz gauges.
Discuss energy density and flow for such a wave.

Deduce the frequency $\omega$ of the longitudinal wave in a hot, 
collisionless plasma that propagates transversely to a uniform 
external magnetic field ${\bf B}_0$ in terms of the (electron)
cyclotron frequency,
\begin{equation}
\omega_B = {e B_0 \over m c}\, ,
\label{p1}
\end{equation}
(in Gaussian units),
the (electron) plasma frequency,
\begin{equation}
\omega_P^2 = {4 \pi N e^2 \over m}\, ,
\label{p2}
\end{equation}
and the electron temperature $T$,
where $e > 0$ and $m$ are the charge and mass of the electron, $c$ 
is the speed of light, and $N$ is the electron number density.

For a simplified analysis, you may assume that the positive ions are at rest, that
all electrons have the same transverse velocity $v_\perp = \sqrt{2 K T / m}$, 
where $K$ is Boltzmann's constant, $T$ is the temperature,  
and that the densities of the ions and unperturbed electrons are uniform. 
% and that the magnitude of the  electric field of the wave is small compared 
% that of the external magnetic field.
Then the discussion may proceed from an (approximate) analysis of the
motion of an individual electron to the resulting
polarization density and dielectric constant, \etc

Such waves are called electron Bernstein waves, following their prediction 
via an
analysis based on the Boltzmann transport equation \cite{Bernstein}.
Bernstein waves were first produced in laboratory plasmas in 1964
\cite{Crawford}, following possible detection in the ionosphere in 1963.
They are now being applied  in plasma diagnostics where it is desired to 
propagate waves below the plasma frequency \cite{Jones}.

\section{Solution}

\subsection{General Remarks}

We first verify that Maxwell's equations imply that when an electric field
{\bf E} has no time dependence, then $\nabla \times {\bf E} = 0$.

If $\partial {\bf E} / \partial t = 0$, then the magnetic field {\bf B} obeys  $\partial^2 {\bf B} / \partial t^2 = 0$, 
as follows on taking the time derivative of Faraday's
law, $c\nabla \times {\bf E} = - \partial {\bf B} / \partial t$ in Gaussian
units.  In principle, this is consistent with a magnetic field that varies
linearly with time, ${\bf B}({\bf r},t) = {\bf B}_0({\bf r}) + 
{\bf B}_1({\bf r})t$.  However,
this leads to arbitrarily large magnetic fields at early and late times, and
is excluded on physical grounds.   Hence, $\partial {\bf E} / \partial t = 0$
implies that $\partial {\bf B} / \partial t = 0$ also, and 
$\nabla \times {\bf E} = 0$ according to Faraday's law.

We next consider some general properties of a longitudinal plane
electric wave, taken to have the form
\begin{equation}
{\bf E} = E_x \hat{\bf x} e^{i(kx - \omega t)}.
\label{s124}
\end{equation} 
This obeys $\nabla \times {\bf E} = 0$, and so can be derived from an 
electric potential, namely 
\begin{equation}
{\bf E} = - \nabla V \qquad \mbox{where} \qquad
V = i{E_x \over k} e^{i(kx - \omega t)}. 
\label{s125}
\end{equation} 
The electric wave (\ref{s124}) has no associated magnetic wave, 
since Faraday's law tells us that 
\begin{equation}
0 = \nabla \times {\bf E} 
= - {1 \over c} {\partial {\bf B} \over \partial t}\, ,
\label{s126}
\end{equation}
and any magnetic field in the problem must be static.

It is well known that electromagnetic waves in vacuum are transverse.  
A longitudinal electric wave can only exist in a medium that can support 
a nonzero polarization density {\bf P} (volume density of electric 
dipole moments).  The polarization density implies an
effective charge density $\rho$ given by
\begin{equation}
\rho = - \nabla \cdot {\bf P} 
%= - {\omega_p^2 \over 4 \pi (\omega^2_B - \omega^2)} \nabla \cdot {\bf E},
\label{s127}
\end{equation}
which is consistent with the first Maxwell equation, 
\begin{equation}
\nabla \cdot {\bf E} = 4 \pi \rho,
\label{s128}
\end{equation}
only if 
\begin{equation}
{\bf P} = - {{\bf E} \over 4 \pi}, 
\label{s129}
\end{equation}
in which case the electric displacement {\bf D} of the longitudinal wave vanishes,
\begin{equation}
{\bf D} = {\bf E} + 4 \pi {\bf P} = 0. 
\label{s130}
\end{equation}
Hence, the (relative) dielectric constant $\epsilon$
%, and the index of refraction$n = \sqrt{\epsilon}$, 
also vanishes  
%This suggests that the longitudinal waves do not actually propagate.

Strictly speaking, eq.~(\ref{s129}) could read ${\bf P} = - {\bf E} / 4 \pi
+ {\bf P}'$, for any field ${\bf P}'$ that obeys $\nabla \cdot {\bf P}' = 0$.
However, since any magnetic field in the problem is static, the fourth Maxwell 
equation tells us that
\begin{equation}
\nabla \times {\bf B} = {4 \pi \over c} \left( {\bf J} + {1 \over 4 \pi}
{\partial {\bf E} \over \partial t} \right)\, 
\label{s131}
\end{equation}
has no time dependence.  Recalling that the polarization current is related by 
\begin{equation}
{\bf J} = {\partial {\bf P} \over \partial t}\, ,
\label{s132}
\end{equation}
we again find relation (\ref{s129}) with the possible addition of a static field
${\bf P}'$ that is associated with a truly electrostatic field ${\bf E}'$.
In sum, a longitudinal electric wave described by eqs.~(\ref{s124}),
(\ref{s129}) and (\ref{s130}) can coexist with background electrostatic and
magnetostatic fields of the usual type.

Maxwell's equations alone provide no relation between the wave number $k$ 
and the wave 
frequency $\omega$ of the longitudinal wave, 
and hence the wave phase velocity $\omega / k$ is arbitrary.
This suggests that purely longitudinal electric waves are best considered
as limiting cases of more general waves, for which additional physical 
relations provide
additional information as to the character of the waves.

\subsection{Gauge Invariance}

Since
the electric wave (\ref{s124}) has no associated magnetic field, we can define its
vector potential {\bf A} to be zero, which is certainly consistent with the
Coulomb gauge condition $\nabla \cdot {\bf A} = 0$.  Suppose, however, we prefer
to work in the Lorentz gauge, for which
\begin{equation}
\nabla \cdot {\bf A} = - {1 \over c} {\partial V \over \partial t}.
% = {\omega E_x \over k c} e^{i(k x - \omega t)}.
\label{s301}
\end{equation}
Then, the vector potential will be nonzero, and the electric field is related by
\begin{equation}
{\bf E} = - \nabla V - {1 \over c} {\partial {\bf A} \over \partial t}
= E_x \hat{\bf x} e^{i(kx - \omega t)}.
\label{s302}
\end{equation}
Clearly the potentials have the forms
\begin{equation}
{\bf A} = A_x \hat{\bf x} e^{i(kx - \omega t)},
\qquad
V = V_0 e^{i(kx - \omega t)},
\label{s303}
\end{equation}
which are consistent with ${\bf B} = \nabla \times {\bf A} = 0$.
From the Lorentz gauge condition (\ref{s301}) we have
\begin{equation}
k A_x = {\omega \over c} V_0,
\label{s304}
\end{equation}
and from eq.~(\ref{s302}) we find
\begin{equation}
E_x = i k V_0 + i {\omega \over c} A_x.
\label{s305}
\end{equation}
Hence,
\begin{equation}
{\bf A} = - i {\omega c \over \omega^2 + k^2 c^2} E_x \hat{\bf x} e^{i(kx - \omega t)},
\qquad
V = - i {k c^2 \over \omega^2 + k^2 c^2} E_x e^{i(kx - \omega t)}.
\label{s306}
\end{equation}

We could also derive the wave (\ref{s124}) from the potentials
\begin{equation}
{\bf A} = - i {c \over \omega} E_x \hat{\bf x} e^{i(kx - \omega t)},
\qquad
V = 0.
\label{s307}
\end{equation}

Thus, an ``electrostatic wave" is not necessarily associated with an
``electrostatic'' scalar potential.

\subsection{Energy Considerations}

A common expression for the electric field energy density is ${\bf E} \cdot {\bf D}
/ 8 \pi$.  However, this vanishes for longitudinal electric waves, according to
eq.~(\ref{s130}).  Further, since the longitudinal electric wave can exist with
zero magnetic field, there is no Poynting vector ${\bf S} = (c / 4 \pi) {\bf E} \times
{\bf H}$ or momentum density ${\bf p}_{\rm field} = {\bf D} \times {\bf B} / 4 \pi c$,
according to the usual prescriptions.

Let us recall the origins of the standard lore.  Namely, the rate of work done by the
field {\bf E} on current density {\bf J} is
\begin{equation}
{\bf J} \cdot {\bf E} = {\partial {\bf P} \over \partial t} \cdot {\bf E}
= - {1 \over 4 \pi} {\partial {\bf E} \over \partial t} \cdot {\bf E}
= - {\partial E^2 / 8 \pi \over \partial t}\, ,
\label{s401}
\end{equation}
using eqs.~(\ref{s129}) and (\ref{s132}).  This work is done at the expense of the
electric field energy density $u_{\rm field}$, which we therefore identify as
\begin{equation}
u_{\rm field} = {E^2 \over 8 \pi}
= {E_x^2 \over 8 \pi} \cos^2 (k x - \omega t),
\label{s402}
\end{equation}
for the longitudinal wave (\ref{s124}).  We readily interpret this energy density
as moving in the $+x$ direction at the phase velocity $v_p = \omega / k$, even
though the derivation of eq.~(\ref{s401}) did not lead to a Poynting vector.

We should also note that energy is stored in the medium in the form of kinetic
energy of the electrons (and, in general, ions as well) that contribute to the 
polarization,
\begin{equation}
{\bf P} = N e ( {\bf x} - {\bf x}_0) = - {{\bf E} \over 4 \pi}\, .
\label{s403}
\end{equation}
Thus, the velocity of an electron is given by
\begin{equation}
{\bf v} = {\bf v}_0 - {\dot {\bf E} \over 4 \pi N e}
= {\bf v}_0 - {\omega E_x \hat{\bf x} \over 4 \pi N e} \sin(k x - \omega t).
\label{s404}
\end{equation}
In squaring this to get the kinetic energy, we neglect the term in
${\bf v}_0 \cdot \hat{\bf x}$, assuming its average to be zero as holds for a medium that is
at rest on average  (and also holds for a plasma in
a tokamak when $x$ is taken as the radial coordinate in a small volume).
  Then, we find the mechanical energy density to be
\begin{equation}
u_{\rm mech} = {1 \over 2} N m v^2 = {1 \over 2} N m v_0^2
+  {E_x^2 \over 8 \pi} {\omega^2 m  \over 4 \pi N e^2} \sin^2(k x - \omega t)
= u_{\rm mech,0}
+  {\omega^2 \over \omega_P^2} {E_x^2 \over 8 \pi} \sin^2(k x - \omega t),
\label{s405}
\end{equation}
where $\omega_P$ is the (electron) plasma frequency (\ref{p2}).
We again can interpret the additional term as an energy density that flows in the
$+x$ direction at the phase velocity.

The total, time-averaged energy density associated with the longitudinal wave is
\begin{equation}
\ave{u_{\rm wave}} = 
{\omega^2 + \omega_P^2 \over  2 \omega_P^2} {E_x^2 \over 8 \pi}\, .
\label{s406}
\end{equation}

If the wave frequency is less than the plasma frequency, as is the case for
examples of Bernstein waves discussed in the sec.~2.5, the longitudinal electric
field energy density is larger than that of the mechanical energy density of the
wave.

\subsection{Longitudinal Waves in a Cold, Unmagnetized Plasma}

As a preliminary exercise we consider the case of a longitudinal wave,
\begin{equation}
{\bf E} = E_x \hat{\bf x}  %+ E_y \hat{\bf y} + E_z \hat{\bf z})
 \cos(k x - \omega t),
\label{s501}
\end{equation}
in a cold, unmagnetized plasma.  An electron at ${\bf x}_0$ in the
absence of the wave has coordinate ${\bf x} = {\bf x}_0 + \delta{\bf x}$
when in the wave, where only the  $x$ component of the equation of motion is
nontrivial:
\begin{equation}
m \delta \ddot x = - e E_x \cos(kx - \omega t)
\approx - e E_x \cos(kx_0 - \omega t).
\label{s502}
\end{equation}
The approximation in eq.~(\ref{s502}) is that the oscillations are small.
Then we find,
\begin{equation}
\delta x \approx {e \over m \omega^2} E_x \cos(kx_0 - \omega t).
\label{s503}
\end{equation}
The resulting electric dipole moment density {\bf P} is
\begin{equation}
{\bf P} = - N e \delta x\ \hat{\bf x} = -{N e^2 \over m \omega^2} {\bf E}
= - {\omega_P^2 \over 4 \pi \omega^2} {\bf E},
\label{s504}
\end{equation}
where $\omega_P$ is the (electron) plasma frequency (\ref{p2}).

For a longitudinal wave, the electric displacement must vanish according to 
eq.~(\ref{s130}), so we find
\begin{equation}
0 = {\bf D} = {\bf E} + 4 \pi {\bf P} 
= \left( 1 - {\omega_P^2 \over \omega^2} \right){\bf E},
\label{s505}
\end{equation}
which requires that
\begin{equation}
\omega = \omega_P.
\label{s506}
\end{equation}
That is, the frequency of longitudinal electric waves can only be the plasma
frequency in a cold, unmagnetized plasma.

\subsection{Longitudinal Waves in a Hot, Magnetized Plasma}

Turning now to the problem of plane waves in a magnetized plasma, we consider waves
whose propagation vector {\bf k} is transverse to the external magnetic field ${\bf B}_0$,
and seek a solution where electric field vector {\bf E} is parallel to {\bf k}.

%The physical picture is that when the oscillatory electric field {\bf E} is perpendicular to
%the static magnetic field ${\bf B}_0$, and the wave frequency is at the cyclotron
%frequency (or harmonics thereof), the cyclotron
%motion of the plasma electrons is perturbed such that the charge density oscillates in sheets
%normal to {\bf E}, and hence are associated with longitudinal
%waves.

We adopt a rectangular coordinate system in which the external magnetic
field ${\bf B}_0$ is along the $+z$ axis and the plane electric wave propagates along 
the $+x$ axis:
\begin{equation}
{\bf E} = E_x \hat{\bf x}  %+ E_y \hat{\bf y} + E_z \hat{\bf z})
 \cos(k x - \omega t).
\label{s1}
\end{equation}
The unperturbed ($E = 0$) motion of an electron is on a helix of radius
\begin{equation}
r_B = {v_\perp \over \omega_B}\, ,
\label{s2}
\end{equation}
where $v_\perp = \sqrt{2 K T / m}$ for all electrons in our simplified analysis.
Hence, we can write the general (nonrelativistic) motion as
\begin{eqnarray}
x & = & x_0 + r_B \cos(\omega_B t + \phi_0) + \delta x,
\label{s3} \\
y & = & y_0 + r_B \sin(\omega_B t + \phi_0) + \delta y,
\label{s4} \\
z & = & z_0 + v_z t + \delta z,
\label{s5}
\end{eqnarray}
noting that the circular motion of a negatively charged electron is counterclockwise
in the $x$-$y$ plane for an external magnetic field along the $+z$ axis.  For an electron
in the collisionless plasma, we consider the Lorentz
force only from the wave electric field and the external magnetic field, $ - e ({\bf E}
+ {\bf v} / c \times {\bf B}_0)$.  The equations of motion are then
\begin{eqnarray}
m [- \omega^2 r_B \cos(\omega_B t + \phi_0) + \delta \ddot x] & = & 
- e E_x \cos(k x - \omega t)  - {e B_0 \over c} [ \omega_B r_B \cos(\omega_B t + \phi_0) 
+ \delta \dot y]
\label{s6} \\
m [- \omega^2 r_B \sin(\omega_B t + \phi_0) +\delta \ddot y] & = & 
- {e B_0 \over c} [ \omega_B r_B \sin(\omega_B t + \phi_0) 
- \delta \dot x]
\label{s7} \\
m \delta \ddot z & = & 0.
\label{s8}
\end{eqnarray}
Recalling eq.~(\ref{p1}) for the cyclotron frequency, the equations of motion reduce to
\begin{eqnarray}
\delta \ddot x + \omega_B \delta \dot y & = & 
- {e E_x \over m} \cos(k x - \omega t),
\label{s9} \\
\delta \ddot y - \omega_B \delta \dot x & = & 
0, 
\label{s10} \\
\delta \ddot z & = & 0.
\label{s11}
\end{eqnarray}
Equation~(\ref{s11}) has the trivial
solution $\delta z = 0$, while eq.~(\ref{s10}) integrates to
\begin{equation}
\delta \dot y = \omega_B \delta x.
\label{s212}
\end{equation}
With this, the remaining equation of motion becomes
\begin{equation}
\delta \ddot x + \omega_B^2 \delta x = - {e E_x \over m} \cos(k x - \omega t),
\label{s213}
\end{equation}
To proceed, we must expand the factor $\cos(k x - \omega t)$, which we do as follows:
\begin{eqnarray}
\cos(k x - \omega t) & = & \cos(k x_0 - \omega t + k r_B \cos(\omega_B t + \phi_0)
+ k \delta x)
\nonumber \\
& \approx & \cos(k x_0 - \omega t + k r_B \cos(\omega_B t + \phi_0))
\nonumber \\
& \approx & \cos(k x_0 - \omega t) \cos(k r_B \cos(\omega_B t + \phi_0))
- k r_B \cos(\omega_B t + \phi_0) \sin (k x_0 - \omega t) 
\nonumber \\
& \approx & \cos(k x_0 - \omega t) \left( 1 - {1 \over 2} k^2 r_B^2 
\cos^2(\omega_B t + \phi_0) \right)
 \nonumber \\
& \approx & \cos(k x_0 - \omega t) \left( 1 - {1 \over 4} k^2 r_B^2 \right)
= \cos(k x_0 - \omega t) \left( 1 - {k^2 v_\perp^2 \over 4 \omega_B^2} \right).
\label{s12}
\end{eqnarray}
In the above we have supposed that $\delta x \ll r_B$ in going from the first line to
the second, that $r_B \ll x_0$ in going from the second line to the third,
that $k r_B \ll 1$ and $\ave{\cos(\omega_B t + \phi_0) \sin (k x_0 - \omega t)}$ = 0
in going from the third line to the fourth, and that 
$\ave{\cos^2(\omega_B t + \phi_0)} = 1/2$ in going from the fourth line to the
fifth.  Perhaps the most doubtful assumption is that $k r_B \ll 1$.

The approximate equations of motion is now
\begin{equation}
\delta \ddot x + \omega_B^2 \delta x = - {e E_x \over m} 
\left( 1 - {k^2 v_\perp^2 \over 4 \omega_B^2} \right) \cos(k x_0 - \omega t).
\label{s13}
\end{equation}
The solution to this is
\begin{equation}
\delta x = - {e \over m (\omega_B^2 - \omega^2)} 
\left( 1 - {k^2 v_\perp^2 \over 4 \omega_B^2} \right) E_x \cos(k x_0 - \omega t).
\label{s14}
\end{equation}
The resulting electric dipole moment density {\bf P} is
\begin{equation}
{\bf P} = - N e \delta x\ \hat{\bf x} =  {N e^2 \over m (\omega_B^2 - \omega^2)} 
\left( 1 - {k^2 v_\perp^2 \over 4 \omega_B^2} \right) {\bf E}
= {\omega_P^2 \over 4 \pi (\omega_B^2 - \omega^2)} 
\left( 1 - {k^2 v_\perp^2 \over 4 \omega_B^2} \right){\bf E},
\label{s15}
\end{equation}
where $\omega_P$ is the (electron) plasma frequency (\ref{p2}).

For a longitudinal wave, the electric displacement must vanish according to 
eq.~(\ref{s130}), so we find
\begin{equation}
0 = {\bf D} = {\bf E} + 4 \pi {\bf P} 
= \left[ 1 + {\omega_P^2 \over \omega_B^2 - \omega^2} 
\left( 1 - {k^2 v_\perp^2 \over 4 \omega_B^2} \right) \right] {\bf E},
\label{s16}
\end{equation}
which requires that
\begin{equation}
\omega^2 = 
\omega_B^2 + \omega_P^2 \left( 1 - {k^2 v_\perp^2 \over 4 \omega_B^2} \right)
= \omega_B^2 + \omega_P^2 \left( 1 - {k^2 K T \over 2 m \omega_B^2} \right).
\label{s17}
\end{equation}
This result corresponds to keeping only the first term of Bernstein's series
expansion, eq.~(50) of \cite{Bernstein}.

In the limit of a cold plasma, where $v_\perp = 0$, the frequency of the
longitudinal wave is $\sqrt{\omega_B^2 + \omega_P^2}$, which is the so-called
upper hybrid resonance frequency.  (This result is well-known to follow from
the assumption of a cold plasma.)

In our model, the effect of nonzero temperature is to lower the frequency of the
longitudinal wave, bringing it closer to the cyclotron frequency, $\omega_B$. 
The effect is greater for shorter wavelengths (larger wave number $k$).  Our
approximation implies that for wavelengths small compared to $r_\perp$, the
characteristic radius of the electron cyclotron motion at temperature $T$,
the frequency of the wave approaches zero.  However, our approximation becomes
doubtful for $k r_\perp \gg 1$.  Bernstein finds that the wave frequency is
restricted to a band around $\omega_B$, which result is only hinted at by
our analysis. 

If we evaluate the dispersion relation (\ref{s15}) at the cyclotron frequency,
$\omega = \omega_B$, then we find the following representative values for
parameters of a Bernstein wave:
\begin{equation}
k = {2 \omega_B \over v_\perp} = {2 \over r_\perp},
\qquad
\lambda = \pi r_\perp,
\qquad \mbox{and} \qquad
v_p = {\omega_B \over k} = {v_\perp \over 2} \ll c.
\label{s19}
\end{equation}

While our analysis does not constrain the phase velocity, $v_p = \omega / k$, of
the longitudinal wave, we do find a relation between $v_p$ and the group velocity,
$v_g = d \omega / dk$:
\begin{equation}
v_g = {d \omega \over dk} = - {\omega_P^2 \over \omega_B^2} {K T \over 2 m v_p}\, .
\label{s18}
\end{equation}
The longitudinal electric waves are negative group velocity waves!  We have written
elsewhere on a paradox associated with this latter phenomenon \cite{negative},
where we found that a negative group velocity can have any magnitude without
contradicting the insight of Einstein that signals must propagate at velocities
less than or equal to $c$.  Hence, the lack of a constraint on $v_p$ is not a
fundamental flaw in the analysis.

\bigskip

The author thanks Brent Jones for bringing this problem to his attention.  For a related discussion of magnetostatic waves, see \cite{spinwave}.

\end{document}